# Negative permittivity attests to local attractive interactions in bubble and stripe phases

B. Friess[1], Y. Peng[2], B. Rosenow[3], F. von Oppen[2], V. Umansky[4], K. von Klitzing[1] and J. H. Smet[1]

**Affiliations:**

[1] Max Planck Institute for Solid State Research, Heisenbergstrasse 1, D-70569 Stuttgart, Germany

[2] Dahlem Center for Complex Quantum Systems and Fachbereich Physik, Freie Universität Berlin, D-14195 Berlin, Germany

[3] Institute for Theoretical Physics, Universität Leipzig, Vor dem Hospitaltore 1, D-04103 Leipzig, Germany

[4] Braun Centre for Semiconductor Research, Department of Condensed Matter Physics, Weizmann Institute of Science, Rehovot 76100, Israel

*Correspondence to: j.smet@fkf.mpg.de

**The physics of itinerant electrons in condensed matter is by and large governed by repulsive Coulomb forces. However, rare cases exist where local attractive interactions emerge and prevail in determining the ground state of the system despite the dominant Coulomb repulsion. The most notable example is no doubt electron pairing, which leads to superconductivity and is mediated by electron phonon coupling or more intricate mechanisms such as antiferromagnetic spin order in high-temperature superconductors[1]. The interplay of attractive and repulsive interaction components may also instigate spontaneous symmetry lowering and clustering of charges in geometric patterns such as bubbles and stripes, provided these interactions act on different length scales. In high-temperature superconductors, for instance, fluctuating stripe or nematic ordering is intertwined with superconductivity itself[2-4]. Both types of attractive interaction based physics, pairing and charge ordering, are also at play in high-quality two-dimensional electron systems exposed to a quantizing perpendicular magnetic field for electrons describing larger orbits. This has been concluded indirectly from the transport behaviour when such phases form. An unambiguous manifestation of an underlying attractive interaction between the electrons has however remained elusive. Here we report the observation of a negative permittivity as an immediate consequence of local attractive interaction rather than repulsion among the electrons when bubble and stripe phases form. The implemented technique based on surface acoustic waves offers directionality and confirms the stripe phase to be a strongly anisotropic medium.**

In a two-dimensional electron system (2DES) exposed to a perpendicular magnetic field, the electrons' kinetic energy is quantized and the electronic states condense onto a set of discrete Landau levels. They are characterized by their orbital index $N = 0, 1, 2,...$ and are each capable of accommodating a number of electrons equal to the number of elementary flux quanta that thread the sample. This macroscopic degeneracy makes the system particularly susceptible to subtle details of the interactions. When the lowest Landau level is partially filled, the direct Coulomb repulsion dominates and brings about the celebrated fractional quantum Hall effect at odd-denominator rational fillings (ν) of a spin resolved Landau level[5]. Electrons occupying higher Landau levels can be described by ring-shaped wave functions whose diameter increases with the orbital index. This expansion affords more room for attractive interactions, such as the exchange term in the Hamiltonian, to influence the ground state despite the pervasive Coulomb repulsion. Ground states driven by local attractive interactions in the 2DES come in two flavours not unlike in high-temperature superconducting materials. They are either based on pairing and Bose-Einstein condensation or the clustering of like charges. At half filling of either spin branch of the lowest Landau level (ν = 1/2 or 3/2), Coulomb repulsion still imposes metallic behaviour which can be elegantly described in terms of weakly interacting composite fermion quasi-particles[6]. They no longer experience the external magnetic field, and their formation amounts to screening of the Coulomb repulsion. For the second Landau level, however, converting the system from strongly interacting electrons to a system of weakly interacting composite fermions with the identical transformation overscreens the Coulomb repulsion. The residual attractive interaction allows for pairing of composite fermions, as evidenced by unconventional fractional quantum Hall physics at even denominator fractional filling 5/2 and 7/2 in GaAs heterostructures hosting the highest quality 2D electron systems, which is considered a manifestation of p-wave superconductivity[7-9]. Seemingly minor alterations of the density and energetics, for instance adding an in-plane magnetic field[10,11], deviating slightly from half filling[12] or applying hydrostatic pressure[13], causes local attractive interactions to evoke distinct ground states characterized by a spatial reordering of the electrons into patterns resembling either bubbles or stripes. Thermal and quantum fluctuations affect the strict periodicity of these charge density waves and generate several distinct phases which can be classified following liquid crystal theory[14]. In higher Landau levels ($N > 1$, $ν > 4$) these charge density wave phases proliferate spontaneously, leaving no room for pairing and even superseding all conventional fractional quantum Hall physics[15-18].

So far, this charge ordering has been primarily studied in macroscopic transport experiments[10,11,19-21]. The bubble phase shows up as reentrant quantum Hall behaviour corresponding to the nearest regular integer quantum Hall state: The crystal of bubbles composed of electrons or holes floats in an insulating, incompressible sea with the nearest integer filling and gets pinned by disorder (Fig. 1a). Near half fillings, anisotropic transport along the two orthogonal principal directions of the GaAs crystal heralds the formation of a stripe phase composed of alternating regions with filling factors of the two nearby integer quantum Hall states (Fig. 2b). For current injection aligned with the orientation of the stripes, the resistance is low but non-zero as current flow benefits from extended states running parallel to stripes with integer filling. While the magneto-transport behaviour is compatible with this charge ordering, it can neither deliver microscopic information about the spatial modulation of the density nor prove the existence of local attractive interactions as the fundamental cause of this symmetry breaking effect. Some efforts to extract microscopic information have been reported in recent years with different techniques[22,23]. However, collecting evidence for an underlying local attractive interaction among the electrons has remained an unaccomplished experimental mission.

The counterintuitive attraction among itinerant electrons can translate into an unconventional response to an applied electric field. Its amplitude would not merely be reduced by screening, but rather its orientation could get reversed and, as a result of this so-called overscreening, the electric permittivity of the two-dimensional electron system would take on a negative value. It is possible to probe this permittivity for an in-plane electric field near the surface, where the two dimensional electron system is located, by exploiting the coupling between the electrical and mechanical properties of the GaAs host crystal. The excitation of an acoustic surface wave is accompanied by a piezoelectric field. This field represents an additional restoring force. It effectively stiffens the lattice and enhances the propagation velocity. The latter is no longer only determined by the mass density $\rho$ and the elastic constant $c$ according to $v_0 = \sqrt{c/\rho}$ but also by the electric permittivity which accounts for this electric restoring force[24-26]. If the GaAs crystal accommodates a 2DES, the piezoelectric field may be partially screened, and the screening capability of the 2D electrons will depend on their ground state. Hence, detecting the propagation velocity of the surface acoustic wave (SAW) discloses the screening behaviour of the electrons. The relative change of the propagation velocity is proportional to the real part of the inverse of the permittivity $\varepsilon$ of the 2DES:

$$\Delta v/v_0 = (v - v_0)/v_0 = K \cdot \text{Re}(1/\varepsilon), \qquad \text{(Eq. 1)}$$

where $\varepsilon$ depends on the frequency and the wave vector of the elastic wave. Here, $v_0$ is the reference propagation velocity for a perfectly conducting system with complete screening of the piezoelectric field in the plane of the two-dimensional electron system. The proportionality constant, $K$, is referred to as the effective coupling constant. It depends solely on material parameters of the host and for GaAs has been extracted from experimental data: $K = 3.2 \cdot 10^{-4}$ (25,26). Further details of how the velocity shift, typically measured in experiment, is related to the electronic properties of the two-dimensional electron system are contained in the supplementary information. This SAW technique is contactless, offers true directionality, and works for insulating and conducting ground states alike. While transport in strong magnetic fields may depend on contact arrangements and an inhomogeneous, field dependent current distribution, SAWs probe the bulk properties of the 2D system in a well-defined area and for a controllable direction.

To perform the SAW experiments, comb-shaped metallic gate structures were fabricated on top of the GaAs substrate as illustrated in Fig. 1c. The fabrication details are deferred to the method section.

Four of these interdigital transducers were placed symmetrically around the square-shaped mesa demarcating the 2DES. When applying an alternating voltage between the two arms of a single transducer, a surface acoustic wave is excited and propagates in the direction perpendicular to the interdigital finger pattern. Due to the piezoelectricity, the acoustic wave is accompanied by a longitudinal electric wave, i.e. the electric field is parallel to the propagation direction. Having reached the transducer at the opposite end, the sound wave is converted back into an electrical signal. It is detected by the network analyser, which can extract the magnetic field dependent SAW velocity shift $\Delta v$ caused by the interaction of the 2DES with the piezoelectric field. An example of the velocity shift recorded as a function of the perpendicular magnetic field is plotted in Fig. 2 and compared with the behaviour of the magnetotransport quantities $R_{xx}$ and $R_H$, the longitudinal and Hall resistances. All measurements were performed along the two principal crystal directions [110] and [1$\bar{1}$0] for easy identification of the anisotropic stripe phase. The stripe phase causes a strong anisotropy in the longitudinal resistance at half-filling of higher Landau levels ($N \geq 2$), whereas the pinned bubble phase manifests itself in a reappearance of the integer quantum Hall plateau outside of the standard series of plateaus. Both phases become more pronounced in transport the lower the Landau level index $N$, consistent with previous publications[19-20].

Analysing the corresponding SAW measurement in Fig. 2b, one recognizes an anti-correlation between the propagation velocity of the surface acoustic wave and the longitudinal sample conductivity, which away from zero field is essentially proportional to $R_{xx}$ due to the tensorial nature of the conductivity. Whenever the 2D electron system becomes less conductive and incompressible in the bulk, i.e. where Shubnikov de Haas oscillations exhibit minima and where quantum Hall states develop, a velocity increase appears. The velocity drops back to its value at low magnetic field when the conductivity increases again. Notable exceptions to this rule are apparent precisely when bubble or stripe phases arise in transport.

The general behaviour outside of the bubble and stripe phase regime is captured well by the relaxation model for SAW propagation developed in Ref. 24 and references therein. It ignores any reactive and frequency dependent components in Eq. 1 and reduces the response of the 2D electron layer to a relaxation process during which the piezoelectric field created by the SAW is screened on a time scale set by the magnetic field dependent dc-conductivity only. Two extreme cases can be considered. For electrons with a large dc-conductivity, the piezoelectric field in the plane of the 2DES is screened entirely and the material behaves as if it were non-piezoelectric. The propagation velocity equals the reference value $v_0$. At zero and low magnetic fields, the electrons are highly mobile and this case applies. All changes of the SAW velocity due to different screening of the piezoelectric field are measured and discussed relative to the velocity $v_0$. Shifts of the velocity to higher values occurring at higher magnetic fields can be interpreted as a partial failure to screen the piezoelectric field. For an insulating piezoelectric material with a dc-conductivity approaching zero, the electric field is left unaffected, and the propagation velocity of the surface acoustic wave is maximally enhanced: $\Delta v/v = K$. For intermediate values of the conductivity σ, the propagation velocity obeys the following expression: $\Delta v/v = K/(1+\sigma^2/\sigma_m^2)$. It describes a non-linear but monotonic rise of the velocity with decreasing conductivity which is measured in units of $\sigma_m$, solely determined by the stiffness coefficient $K$ and the average static permittivity of vacuum and the sample half space. Upon entering the quantum Hall regime, the bulk turns more and more incompressible, i.e. insulating, as we approach the quantum Hall filling factor. Accordingly, the propagation velocity keeps increasing in the experiment of Fig. 2 in qualitative agreement with the relaxation model.

In the bubble and stripe regime, however, the propagation velocity displays very different behaviour. Most striking is a negative velocity shift for both propagation directions in the bubble phase region and for parallel propagation in the stripe phase region. The relaxation model utterly fails to account for a negative velocity shift. Instead, Eq. 1 implies that the permittivity of the 2DES turns negative. Using Eq. 1 in conjunction with the effective coupling constant extracted in the literature yields negative permittivity values as large as $1/\varepsilon = -0.2$. In the bubble and stripe regime, the 2D electrons apparently overscreen the piezoelectric field so as to generate a net field with opposite sign that locally enhances the lattice deformation rather than acting as a restoring force opposing the deformation. To the best of our knowledge, such a lattice softening and the accompanying drop in the propagation velocity has not been reported previously in SAW experiments. A negative permittivity may arise near and above a resonance frequency of the electronic system in analogy with the classical behaviour of an externally driven Lorentz oscillator. For the present system, a plausible electronic resonance limited to the bubble and stripe regime would be a pinning mode resonance due to the motion of charge density wave domains pinned in the disorder potential of the host crystal. Such pinning mode resonances have been explored extensively in microwave absorption experiments using coplanar waveguide geometries[27,28]. Note that these microwaves interrogate the dynamical conductivity of the system at effectively zero wave vector, whereas the SAW measures the electromagnetic response not only at finite frequency but also finite wave vector due to the large disparity between the velocity of the microwaves and the SAW (five orders of magnitude). In the literature, the observed microwave absorption resonances range from 200-500 MHz for the bubble phase[27] and 100-150 MHz for the stripe phase[28]. The outcome of such a microwave experiment for the heterostructure studied here is summarized in Fig. 3. The insert in panel c illustrates the coplanar waveguide geometry. The meander offers the advantage of maximizing the interaction length with the microwaves for improved signal-to-noise ratio but is not suitable for probing the anisotropic stripe phase. The pinning mode resonance for the bubble phases when the $N = 2$ Landau level is quarter filled with holes or electrons exceeds 500 MHz. This is well above the 340 MHz at which the measurements in Fig. 1 have been recorded. Hence, it seems the pinning mode resonance can be safely discarded as the sole cause of the observed negative permittivity. This assertion is further corroborated by repeating the SAW experiment in Fig. 3b at even lower frequency. Fig. 3b displays data recorded for 70 MHz. In addition, we performed measurements at frequencies much higher than the pinning mode resonance (up to 1GHz). For the entire range, a negative velocity shift in the CDW regime persists and an exclusive pinning mode resonance scenario to account for the negative permittivity has definitely become untenable. It must be complemented by local attractive interactions among the electrons as a key ingredient to explain overscreening behaviour across such a wide frequency range. This assertion is confirmed by a detailed theoretical description of the dielectric response of the bubble phase outlined in the supplementary information. It identifies local attractive interactions as a crucial mechanism to account for the negative permittivity observed in experiment. In contrast to the above-mentioned relaxation model, the theory includes reactive effects due to the elastic response, pinning as well as the long range repulsive Coulomb interaction. Dissipation accounting for the width of the pinning resonance does not significantly alter the results. Local attractive interactions enter via the local thermodynamic compressibility χ: $\frac{1}{\chi} = \frac{\partial \mu}{\partial n_{2D}}$. Here, μ is the chemical potential and $n_{2D}$ the electron density. The compressibility becomes negative in the bubble phase regime, i.e. the chemical potential drops with increasing charge carrier density reflecting the tendency of electrons to cluster. Negative compressibility is another manifestation of local attractive interaction and has made appearance for all dimensionalities of strongly-interacting electron systems[29-35]. The supplementary information contains numerical results for the permittivity in the bubble phase regime. When using parameter values appropriate for the experiment, the theory yields negative permittivity values comparable to those extracted from the experiment.

The occurrence of a negative velocity shift is intimately connected with the emergence of either bubble or stripe phases. For instance, when returning to Fig. 2 and inspecting the longitudinal resistance at magnetic fields above 3T corresponding to a partial filling of the second or lowest Landau level, a transport hallmark for either phase is absent and the velocity shift indeed remains positive throughout. In principle the second Landau level renders the whole potpourri of correlated ground states of two dimensional electrons. This includes odd-denominator fractional quantum Hall states, even denominator fractional quantum Hall states, which replace the anisotropic stripe phase at half filling unless an in-plane magnetic field is applied[10,11,23,36,37], as well as reentrant quantum Hall phases reminiscent of the bubble phases in higher Landau levels. These reentrant states, however, do not show up in Fig. 2, because of the sample quality and insufficient cooling of the electron system. It should be noted that their increased fragility and fourfold appearance per spin branch instead of twofold appearance in higher Landau levels, has even sparked a debate whether they really are equal in nature to the bubble phases in higher Landau levels[38]. By employing samples of higher quality and optimizing the high frequency signal transmission, it has been possible to lower the intensity of the SAW at the expense of signal-to-noise ratio for the SAW detection while maintaining the electronic temperature below 20 mK. Under these improved experimental conditions, we are able to observe three of the four known reentrant states in each spin branch of the second Landau level (Fig. 4). A comparison of the transport data with the SAW propagation velocity imposingly confirms a one-to-one correlation between the reentrant quantum Hall behaviour associated with bubble like phases and a negative velocity shift of SAWs. We note that at the temperature of this experiment the bubble phases in higher Landau levels have merged with the nearby integer quantum Hall states making it impossible to infer their existence from transport measurements. The presence of the bubble and stripe phases remains however discernible in the SAW transmission data (Fig. 4b). This highlights the preeminence of the employed technique to study phases that internally, on a mesoscopic scale, remain compressible in nature while being incompressible on the macroscopic scale of a dc-transport experiment.

Finally, we turn our focus to the strong anisotropy of the propagation velocity in the stripe phase regime in Fig. 2b. A negative velocity shift is only observed for propagation of the SAW collinear to the direction of the stripes, i.e. parallel to the so-called 'easy' axis as deduced from the transport data in panel a. For propagation along the 'hard' direction, the velocity shift is noticeably positive instead, indicating that the electrons clustered in stripes struggle in screening a periodic piezoelectric field pointing perpendicular to the stripes. The theory for this more complex case in the stripe phase regime is discussed further in the supplementary information (section V). Ultimately the translational invariance along the direction of the stripes is responsible for this positive velocity shift. Little restoring force exists to antagonize an electric field, which is periodic perpendicular to the stripes but uniform along the stripes. For propagation along the easy direction the same ingredients – pinning, Coulomb interaction and negative compressibility – generate a scenario identical to the bubble phase with negative velocity shift and negative permittivity over a wide frequency span. They again result from the prevailing local attractive interaction among the itinerant electrons.

The long-standing mission of establishing experimentally that local attractive interactions provide the impetus for the formation of spatially periodic patterns of the charge density has been accomplished successfully. The capability of electrons participating in such bubble or stripe ordering to decelerate a surface acoustic wave across a wide span of frequencies by overscreening the accompanying piezoelectric field unveils the existence and significance of these interactions.

## Methods

**Sample details**

All measurements were done on GaAs/AlGaAs heterostructures containing a 30 nm-wide GaAs quantum well at a distance of 160 nm from the surface. Electrons are provided by doping layers placed symmetrically on either side of the quantum well at a distance of about 80 nm. The electron densities vary between $3.1 \times 10^{11}$ cm$^{-2}$ (sample for Fig. 2) and $2.9 \times 10^{11}$ cm$^{-2}$ (sample for Fig. 4). High sample qualities have been achieved by applying a superlattice doping scheme[39]. Low-temperature mobilities are in the range of 20-30×10$^6$ cm$^2$/Vs. Samples were patterned into square-shaped mesa structures of width 1.1 mm using standard photolithography. The mesa step height approximately equals the depth of the 2DES. On either side of the mesa an interdigital transducer was fabricated from aluminium as shown in Fig. 1c. The period of the finger structure determines the wavelength of the SAW. The SAW measurements for Fig. 2 and 4 were done on structures with a period of 8.6 µm, which corresponds to a SAW frequency of about 340 MHz. In order to distinguish the SAW propagation along the two principal crystal axes by their traveling time, the transducers were placed further apart in one direction. After completing all fabrication steps, the sample was cooled in a dilution refrigerator to a base temperature of about 20 mK.

**Resistive transport measurements**

Four-point resistance measurements were used to identify the presence of bubble and stripe phases. An external ac current with an amplitude of 2-10 nA and a frequency between 10-20 Hz was imposed along the two principal crystal directions, and the resulting voltage drops were recorded by heterodyne detection.

**Surface acoustic wave measurements**

The SAW propagation was measured using a network analyser for both excitation as well as detection after traversing the mesa structure. The relative change of the SAW velocity $\Delta v/v_0$ is determined from the detected shift of the SAW phase $\Delta\alpha$: $\Delta v/v_0 = \lambda/L \cdot \Delta\alpha/360°$, where L denotes the interaction length (size of the 2DES) and λ the SAW wavelength. By convention, the velocity $v_0$ corresponds to the case where the 2D electrons completely screen the piezoelectric field in the plane of the electron system, which in the present high-quality GaAs/AlGaAs samples is well approximated by the measured phase shift at low magnetic fields[26,40]. By exploiting the time gating capabilities of our network analyser, we were able to individually address the [110] and the [1$\bar{1}$0] directions by the distinct traveling times of the SAW signal resulting from the different separation of the transducers along these two directions. The excitation power of the SAW was kept sufficiently low to avoid heating and nonlinear effects. The signal-to-noise ratio was improved by averaging over multiple cycles and by using an additional external amplifier (1:10$^4$). The SAW measurements were recorded simultaneously to the resistive transport measurements. For the sake of completeness we note that the SAW data below 100 mT has been discarded due to an abrupt phase change associated with the superconducting transition of the aluminium transducers.

**Acknowledgments**

We acknowledge financial support from the German Israeli Foundation (J.H.S. and V.U.), the Collaborative Research Center 183 (F.v.O.), the Priority Programme 1666 (F.v.O.) and grant RO 2247/8-1 (B.R.) of the Deutsche Forschungsgemeinschaft. We thank A. Wixforth and H. Krenner for helpful discussions.

**Author contributions**

B.F. and J.H.S. conceived and designed the experiment. V.U. developed the sample material. B.F. fabricated the samples, carried out the experiments, and performed the data analysis. All authors contributed to the discussion and interpretation of the results. Y.P., B.R. and F.v.O. performed the theoretical modelling and wrote the supplementary material. B.F. and J.H.S wrote the manuscript with input from all authors.

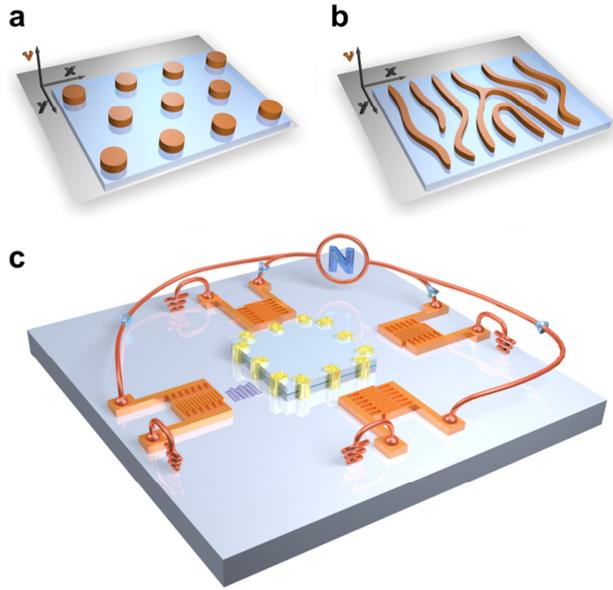

**Fig. 1: Density modulated phases in the quantum Hall regime: Theory and experimental setup**
(a, b) Illustration of the spatial filling factor modulation in the bubble and stripe phase. Instead of a homogeneous density distribution, the electrons in the bubble and stripe phase tend to arrange such that the topmost Landau level is either completely full (indicated in brown) or empty (blue). (c) Experimental setup used to study the interaction of surface acoustic waves and the two-dimensional electron system. The two-dimensional electron system is confined to a square-shaped mesa structure. Four interdigital transducers are placed symmetrically on either side of the mesa. SAWs are generated and detected by a network analyser (N). The SAW propagation velocity depends sensitively on the screening behaviour of the electrons in the 2DES.

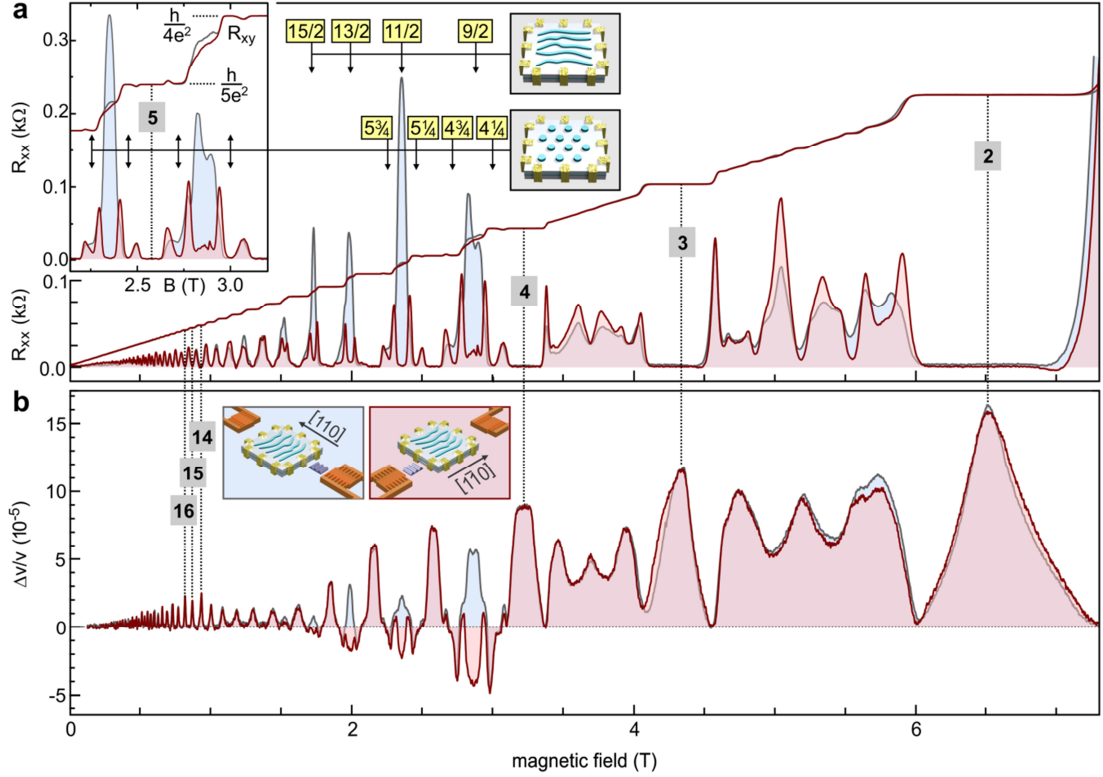

**Fig. 2: Density modulated phases in the quantum Hall regime: Experiment** (a) Longitudinal and Hall resistance for current flow along the principal crystal directions [110] (blue) and [1$\bar{1}$0] (red). The inset shows a magnification of the transport data between filling factor 4 and 6. Bubble and stripe phases are evident by the anisotropic transport behaviour and a reentrance of the integer quantum Hall effect. (b) SAW velocity shift measured simultaneously to the transport data in panel a along the same crystal directions. A negative shift of the SAW velocity is observed in the bubble and stripe phases. The velocity shift is isotropic for the bubble phase but strongly anisotropic for the stripe phase.

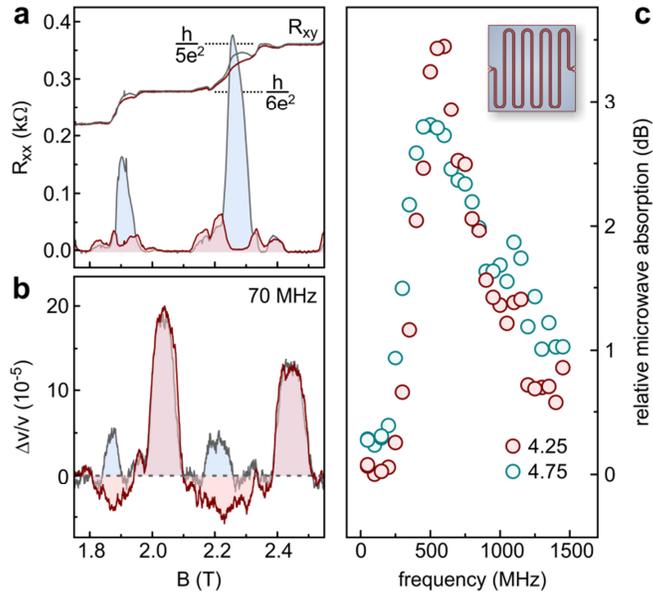

**Fig. 3: On the origin of the negative velocity shift** (a) Longitudinal and Hall resistance between filling factor 5 and 7 in the crystallographic directions [110] (blue) and [1$\bar{1}$0] (red). (b) SAW velocity change measured along the same directions at a SAW frequency of 70 MHz. (c) Frequency dependence of the microwave adsorption at filling factor 4.25 (red circles) and 4.75 (green circles). The measurement was done using a coplanar stripline configuration as shown in the inset. A clear resonance is observed at a frequency of 500-600 MHz.

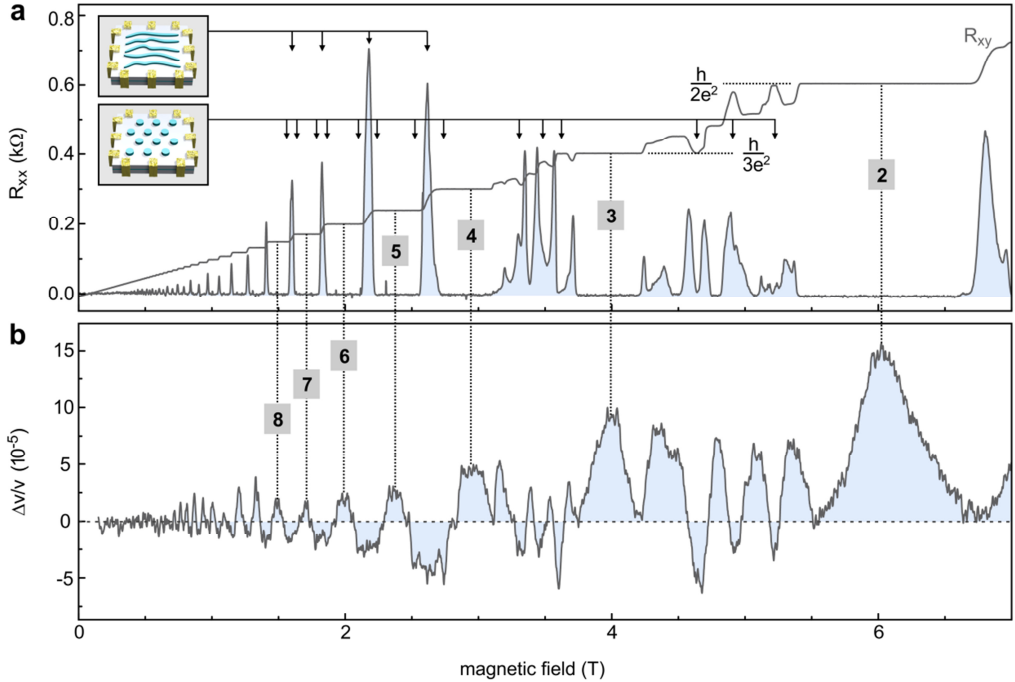

**Fig. 4: Reentrant states in the second Landau level** (a) Longitudinal and Hall resistance measured on a sample with improved quality and lower electron temperature (at the expense of a reduced signal-to-noise ratio for the SAW detection in panel b). Marked are the reentrant phases in the second Landau level as well as the bubble and stripe phases in higher Landau levels. (b) SAW velocity shift measured simultaneously to the transport data in panel a. At low temperatures, a negative velocity change occurs also for the reentrant phases in the second Landau level, supporting their idenfication as bubble phases.